\documentclass[10 pt, letter]{article}
\usepackage{graphicx}
\usepackage{caption}
\usepackage{amsmath}
\usepackage{amsthm}
\usepackage{enumerate}
\usepackage{subfig}
\usepackage{float}
\usepackage{amsfonts}
\usepackage{amssymb}
\usepackage{anysize}
\usepackage{helvet}
\usepackage{setspace}
\usepackage{multirow}
\usepackage{tabularx}
\usepackage{tabulary}
\usepackage{longtable}
\usepackage{float}
\usepackage{mathtools}
 \def\ds{\displaystyle}
\title{\bf Quantization of the 1-D harmonic oscillator with variable mass using the operators $\hat v$ and $\hat p$}
\author{Gustavo V. L\'opez\footnote{gulopez@cencar.udg.mx}~ and Eric M. Reynaga~\\ \\
 Departamento de F\'{i}sica, Universidad de Guadalajara,\\
 Blvd. Marcelino Garc\'{i}a Barragan y Calzada Ol\'{i}mpica, \\Ê44200 Guadalajara, Jalisco, Mexico}
\begin{document}
\maketitle

%\centerline{\large\bf Abstract}
\begin{abstract}
\noindent
For the 1-D harmonic oscillator with position depending variable mass, a Hamiltonian and constant of motion are given through a consistent approach. Then,
the quantization of this system is carried out using the  operator  $\hat p$, for the Hamiltonian,  and the operator $\hat v$ for 
the constant of motion. We find that the result of both  quantizations brings about different quantum dynamics. 
\end{abstract}
\newpage
\section{ Introduction}
Mass variable systems have been important since the beginning of the Classical Mechanics [1-5], and now they are becoming even important in Quantum Mechanics
[6-9]. These type of systems are not invariant under Galileo (non relativistic) or Lorentz (relativistic) transformations [10]. However, Newton's equation of motion can be still be used to study these non relativistic systems [11] as a good approximation of their dynamics. Taking this point of view, a consistent approach for 1-D conservative systems with position depending mass [12] has been already made, 
where an effective potential depending of the variation of mass appears. This effective potential is missing in other approaches [15,16,17] where quantization of these systems is also studied. We consider that this effective potential must be included in any attempt to get a Lagrangian or Hamiltonian for these type of systems, and, of course, it must have a great deal of importance when one is dealing with the quantization of these systems.  In this paper, we will use the effective potential approach to study the quantization of the harmonic oscillator with position depending mass. In addition, we make the comparison of the energy labels resulting from the quantization of the Hamiltonian ( using the $\hat x$ and $\hat p=-i\hbar\partial/\partial x$ operators), and the quantization of the constant of motion ( using the $\hat x$ and $\hat v=-i(\hbar/m)\partial/\partial x$ operators).  \\Ê\\
\section{ Constant of motion and Hamiltonian}
There is an expressions found in [12] for the constant of motion and Hamiltonian of a conservative system with position depending mass,
\begin{subequations}
\begin{equation}
K(x,v)=\frac{m^2(x) v^2}{2m_0}-\frac{1}{m_0}\int_{x_0}^{x}m(s)F(s)~ds
\end{equation}
and
\begin{equation}
H(x,p)=\frac{m_0p^2}{2m^2(x)}-\frac{1}{m_0}\int_{x_0}^{x}m(s)F(s)~ds,
\end{equation}
\end{subequations}
where $m_0=m(x_0)$, and $x_0$ are the mass and the position of the particle at the initial time $t_0$, and $F(x)$ is the external conservative force. Applying these expression for the harmonic oscillator, where the force is given by $F(x)=-kx$ (with $k$ being a constant), the above expressions are written as
\begin{subequations}
\begin{equation}
K(x,v)=\frac{m^2(x) v^2}{2m_0}+\frac{k}{m_0}\int_{x_0}^{x}sm(s)~ds
\end{equation}
and
\begin{equation}
H(x,p)=\frac{m_0p^2}{2m^2(x)}+\frac{k}{m_0}\int_{x_0}^{x}sm(s)~ds,
\end{equation}
\end{subequations} 
To proceed with the analysis, one needs a model for $m(x)$. Let us assume that
\begin{equation}\label{mo1}
m(x)=m_0+m_1x,
\end{equation}
where $m_1=dm(x)/dx$ is the rate of increasing or decreasing the mass of the system. With this model, the constant of motion and Hamiltonian are ($x_0=0$)
\begin{subequations}
\begin{equation}\label{k1}
K(x,v)=\frac{(m_0+m_1x)^2}{2m_0}v^2+\frac{1}{2}m_0\omega^2x^2+\frac{m_1\omega^2}{3}x^3
\end{equation}
and
\begin{equation}\label{h1}
H(x,p)=\frac{m_0p^2}{2(m_0+m_1x)^2}+\frac{1}{2}m_0\omega^2x^2+\frac{m_1\omega^2}{3}x^3,
\end{equation}
\end{subequations} 
where we have defined $\omega=\sqrt{k/m_0}$. Assuming $m_1x/m_0\ll 1$ and using up to second order in Taylor expansion in the Hamiltonian, these expression are 
\begin{subequations}
\begin{equation}\label{k2}
K(x,v)=\frac{1}{2}m_0v^2+\frac{1}{2}m_0\omega^2 x^2+m_1xv^2+\frac{m_1^2}{2m_0}x^2v^2+\frac{m_1\omega^2}{3}x^3
\end{equation}
and
\begin{equation}\label{h2}
H(x,p)=\frac{p^2}{2m_0}+\frac{1}{2}m_0\omega^2x^2+\biggl(-\frac{m_1x}{m_0}+\frac{3m_1^2x^2}{2m_0^2}\biggr)\frac{p^2}{m_0}+\frac{m_1\omega^2}{3}x^3,
\end{equation}
\end{subequations} 
that is, the constant of motion and Hamiltonian are of the form
\begin{equation}\label{KW}
K=K_0+W_K, 
\end{equation}
and
\begin{equation}\label{HW}
 H=H_0+W_H,
\end{equation}
where $K_0$, $W_K$, $H_0$, and $W_H$ are defined as
\begin{subequations}
\begin{equation}\label{K0}
K_0=\frac{1}{2}m_0v^2+\frac{1}{2}m_0\omega^2 x^2,
\end{equation}
\begin{equation}\label{Wk}
W_K=m_1xv^2+\frac{m_1^2}{2m_0}x^2v^2+\frac{m_1\omega^2}{3}x^3\end{equation},
\end{subequations} 
\begin{subequations}
\begin{equation}\label{H0}
H_0=\frac{p^2}{2m_0}+\frac{1}{2}m_0\omega^2x^2,
\end{equation}
and
\begin{equation}\label{Wh}
W_H=\biggl(-\frac{m_1x}{m_0}+\frac{3m_1^2x^2}{2m_0^2}\biggr)\frac{p^2}{m_0}+\frac{m_1\omega^2}{3}x^3.
\end{equation}
\end{subequations} 
\section{Quantization}
To see whether or not the quantum dynamics described by the relations (\ref{KW}) and (\ref{HW}) are different, it is enough to consider $W_K$ and $W_H$ as perturbation of $K_0$ and $H_0$ respectively. The quantization with the Hamiltonian is carried out through the usual association  of an Hermitian operator 
${\widehat H}(\hat x, \hat p)$ [13], being the Hermitian operator $\hat p$ defined as $\hat p=-i\hbar\partial/\partial x$, and solving the Schr\"odinger's equation
\begin{equation}\label{SH}
i\hbar\frac{\partialÊ|\Psi\rangle}{\partial t}={\widehat H}(\hat x, \hat p)|\Psi\rangle.
\end{equation} 
The quantization with the constant of motion is also carried out through the association of an Hermitian operator ${\widehat K}(\hat x, \hat v)$, being the Hermitian operator $\hat v$ defined as $\hat v=-i(\hbar/m_0)\partial/\partial x$, and solving the Schr\"odinger-like equation
\begin{equation}\label{SK}
i\hbar\frac{\partial |\Psi\rangle}{\partial t}={\widehat K}(\hat x,\hat v)|\Psi\rangle.
\end{equation}
Because of the expressions (\ref{k2}) and (\ref{h2}), the equations (\ref{SH}) and (\ref{SK}) corresponds to autonomous systems,  the proposition
\begin{equation}\label{pro}
|\Psi\rangle=e^{-i E t/\hbar}|\Phi\rangle
\end{equation}
reduces the solutions  to solve the eigenvalue problems
\begin{equation}\label{eH}
{\widehat H}|\Phi\rangle_H=E_H|\Phi\rangle_H,
\end{equation}
and
\begin{equation}\label{eK}
{\widehat K}|\Phi\rangle_K=E_K|\Phi\rangle_K.
\end{equation}
Since $W_H$ and $W_K$ are considered as perturbations of the harmonic oscillator with constant mass , it is enough to know the eigenvalues up to second order in perturbation theory to see whether or not there is a difference on the quantum dynamics. From the perturbation theory, it is well known [13] that up to second order on perturbation theory, the eigenvalues are given as
\begin{equation}
E_{\xi,n}=E_{\xi,n}^{(0)}+E_{\xi,n}^{(1)}+E_{\xi,n}^{(2)},\quad\quad \xi=H,K,
\end{equation}
with $E_{\xi,n}^{(0)}$ being the eigenvalues  associated to $H_0$ or $K_0$,
\begin{equation}\label{e0}
E_n^{(0)}=E_{H,n}^{(0)}=E_{K,n}^{(0)}=\hbar\omega(n+1/2).\quad\quad n\in{\cal Z^+},
\end{equation} 
and $E_{\xi,n}^{(1)}$ and $E_{\xi,n}^{(2)}$ are given by
\begin{equation}\label{app1}
E_{\xi,n}^{(1)}=\langle n|\widehat W_{\xi}|n\rangle
\end{equation}
and
\begin{equation}\label{app2}
E_{\xi,n}^{(2)}=\sum_{m\not=n}\frac{\displaystyle|\langle m|\widehat W_{\xi}|n\rangle|^2}{E_n^{(0)}-E_m^{(0)}}.
\end{equation}
The eigenstates $\{|n\rangle\}$ are the eigenstates of $H_0$ or $K_0$ which represents the functions
\begin{equation}
\Phi_n(x)=\langle x|n\rangle=c_ne^{-\alpha^2x^2/2}H_n(\alpha x), \quad \quad c_n=\sqrt{\frac{\alpha}{\sqrt{\pi}~2^nn!}},
\end{equation}
being $H_n$ represents the Hermit polynomials,  and the constant $\alpha$ is given by $\alpha=\sqrt{m\omega/\hbar}$. 
\subsection{Eigenvalues of $E_{H,n}$.}
As we can see from (\ref{h2}), one requires to assign Hermitian operators to the functions $xp^2$, $x^2p^2$ and $x^3$. This can be obtained by using Weyl quantization method [14], or identifying the powers of polynomial $(x+p)^l$ with the powers of operator polynomial $(\hat x+\hat p)^l$ for $l=2,4$. Doing either of these approaches, and using the commutation relation $[\hat x, \hat p]=i\hbar I$ (being I the identity operator), one gets
\begin{subequations}
\begin{equation}
\widehat{xp^2}=\hat x\hat p^2-i\hbar\hat p,
\end{equation}
\begin{equation}
\widehat{x^2p^2}=\hat x^2\hat p^2-i2\hbar\hat x\hat p-\frac{\hbar^2}{2}I,
\end{equation}
and
\begin{equation}
\widehat{x^3}=\hat x^3.
\end{equation}
\end{subequations}
In this way, the associated Hermitian operator to the function $W_H$ is
\begin{equation}
\widehat W_H=-\frac{m_1}{m_0^2}\bigl(\hat x \hat p^2-i\hbar \hat p\bigr)+\frac{3m_1^2}{2m_0^3}\bigr(\hat x^2\hat p^2-i2\hbar\hat x\hat p-\frac{\hbar^2}{2}\bigr)+
\frac{m_1\omega^2}{3}\hat x^3.
\end{equation}
for convenience during evaluation of the matrix elements of the perturbation terms, one uses the ascent ($a^{\dagger}$) and descent ($a$) operators instead of $\hat x$ and $\hat p$,
\begin{equation}
\hat x=\sqrt{\frac{\hbar}{2m_0\omega}}~(a+a^{\dagger})\quad\quad\hbox{and}\quad \hat p=-i\sqrt{\frac{m_0\hbar\omega}{2}}~(a-a^{\dagger}), 
\end{equation} 
where one has the following properties
\begin{equation}
a|n\rangle=\sqrt{n}|n-1\rangle, \quad a^{\dagger}|n\rangle=\sqrt{n+1}|n+1\rangle, \quad\hbox{and}\quad [a,a^{\dagger}]=1.
\end{equation}
Thus, one obtains  $\widehat W_H$ in terms of $a$ and $a^{\dagger}$ as
\begin{eqnarray}
\widehat W_H&=&-\frac{\ds m_1\hbar\omega}{\ds 2m_0}\sqrt{\ds \frac{\hbar}{\ds 2m_0\omega}}
\biggl(a^3-a^2a^{\dagger}-aa^{\dagger}a+a(a^{\dagger})^2+a^{\dagger}a^2-a^{\dagger}aa^{\dagger}-(a^{\dagger})^2 a+a^3\biggr)-
\frac{m_1\hbar}{m_0^2}\sqrt{\frac{m_0\hbar\omega}{2}}~\bigl(a-a^{+}\bigr)
\nonumber\\
& &-\frac{\ds 3m_1^2\hbar^2}{8m_0^3}\biggl(a^4-a^2a^{\dagger}a-a^2(a^{\dagger})^2+a^{\dagger}a^3-a^{\dagger}aa^{\dagger}a-a^{\dagger}a^2a^{\dagger}+a^{\dagger}a(a^{\dagger})^2+aa^{\dagger}a^2-a^3a^{\dagger}-(a^{\dagger})^2aa^{\dagger}\nonumber\\
& &\quad\quad\quad\quad\quad-a^2a^{\dagger}a-aa^{\dagger}aa^{\dagger}+a(a^{\dagger})^3+(a^{\dagger})^2a^2-(a^{\dagger})^3 a+(a^{\dagger})^4\biggr)\nonumber\\
& &-\frac{\ds 3m_1^2\hbar^2}{2m_0^3}\biggl(a^2-aa^{\dagger}+a^{\dagger}a-(a^{\dagger})^2+\frac{1}{2}\biggr)\nonumber\\
& &+\frac{\ds m_1\omega^2}{\ds 3}\left(\frac{\ds \hbar}{\ds 2m_0\omega}\right)^{3/2}
\biggl(a^3+aa^{\dagger}a+a^2a^{\dagger}+a (a^{\dagger})^2+a^{\dagger}a^2+(a^{\dagger})^2a+a^{\dagger}aa^{\dagger}+(a^{\dagger})^3\biggr)\label{whh}
\end{eqnarray}
Therefore, using (\ref{pro}), (\ref{whh}) in (\ref{app1}) and (\ref{app2}), it follows that
\begin{subequations}
\begin{equation}
E_{H,n}^{(1)}=\sigma\left\{\frac{2n^2+2n-1}{4}+\frac{1}{2}\right\}
\end{equation}
and
\begin{equation}
E_{H,n}^{(2)}=-\frac{1}{\hbar\omega}\left\{(\eta-m_1\beta)^2(3n^2+3n+2)+(3\eta-m_1\beta)^2(3n^3+3n+1)+\left(\frac{\sigma}{4}\right)^2(4n^3+6n^2+14n+6)\right\},
\end{equation}
\end{subequations}
where the constants $\sigma$, $\beta$, and $\eta$ have been defined as
\begin{equation}
\sigma=\frac{3m_1^2\hbar^2}{2m_0^3},\quad\quad\beta=\frac{\hbar}{2m_0}\sqrt{\frac{\hbar\omega}{2m_0}}, \quad\hbox{and}\quad \eta=\frac{m_1\omega^2}{3}\left(\frac{\hbar}{2m_0\omega}\right)^{3/2}.
\end{equation}
Thus, up to second order in perturbation theory, the energy of the system of the nth-state is
\begin{eqnarray}\label{energyH}
E_{H,n}&=&\hbar\omega(n+1/2)+\sigma\left\{\frac{2n^2+2n-1}{4}+\frac{1}{2}\right\}\nonumber\\
& &-\frac{1}{\hbar\omega}\left\{(\eta-m_1\beta)^2(3n^2+3n+2)+(3\eta-m_1\beta)^2(3n^3+3n+1)+\left(\frac{\sigma}{4}\right)^2(4n^3+6n^2+14n+6)\right\}.\nonumber\\
\end{eqnarray}
\subsection{ Eigenvalues of $E_{K,n}$}
As we can see from expression (\ref{k2}), and from the previous calculation we have made, one needs to assign Hermitian operators to the functions $xv^2$, $x^2v^24$ and $x^3$. In this case, one has that
\begin{equation}\label{comv}
\hat v=-i\frac{\hbar}{m_0}\frac{\partial}{\partial x}, \quad \quad\hbox{and}\quad [\hat x,\hat v]=i\frac{\hbar}{m_0}I.
\end{equation} 
Using the same method we used previously and the commutation relation (\ref{comv}), one has the following Hermitian operator for $W_K$
\begin{equation}
\widehat W_K=m_1\biggl(\hat x\hat v^2-i\frac{\hbar}{m_0}\hat v\biggr)+\frac{m_1^2}{2m_0}\biggl(\hat x^2\hat v^2-i\frac{2\hbar}{m_0}\hat x\hat v-\frac{\hbar^2}{2m_0^2}\biggr)+\frac{m_1\omega^2}{3m_0}\hat x^3.
\end{equation}
Now, instead of the operators $\hat x$ and $\hat v$ , one changes to the ascend ($a^{\dagger}$) and descend ($a$) operators through the relations
\begin{equation}
\hat x=\sqrt{\frac{\hbar}{2m_0\omega}}~(a+a^{\dagger}),\quad\quad\hbox{and}\quad \hat v=-i\sqrt{\frac{\hbar\omega}{2m_0}}~(a-a^{\dagger}),
\end{equation}
where $a$ and $a^{\dagger}$ have the same properties written in (\ref{pro}). Proceeding similarly as what we did previously, one gets the following expression 
for $\widehat W_K$
\begin{eqnarray}
\widehat W_K&=&
-\frac{\hbar}{2}\frac{m_1}{m_0}\sqrt{\frac{\hbar\omega}{2m_0}}\biggl(\hat a^3-\hat a^2\hat a^+-\hat a\hat a^+\hat a+\hat a\hat a^{+2} +\hat a^+\hat a^2-\hat a^+\hat a\hat a^+-\hat a^{+2}\hat a+\hat a^{+3}\biggr)-2m_1\beta(\hat a-\hat a^+)\nonumber\\ \nonumber\\
& & -\frac{\alpha}{12}\biggl\{\hat a^4-\hat a^2\hat a^+\hat a-\hat a^2\hat a^{+2}+\hat a^+\hat a^3-\hat a^+\hat a\hat a^+\hat a
-\hat a^+\hat a^2\hat a^++\hat a^+\hat a\hat a^{+2}+\hat a\hat a^+\hat a^2 -\hat a\hat a^{+2}\hat a-a^3a^{\dagger}\nonumber\\
& &\quad\quad\quad\quad\quad-\hat a\hat a^+\hat a\hat a^++\hat a\hat a^{+3} +\hat a^{+2}\hat a^2-\hat a^{+3}\hat a-\hat a^{+2}\hat a\hat a^++\hat a^{+4}\biggr\}\nonumber\\Ê\nonumber\\
	& & -\frac{1}{2}\frac{m_1^2\hbar^2}{m_0^3} \biggl(\hat a^2-\hat a\hat a^+-\hat a^+\hat a+\hat a^{+2}+\frac{1}{2}\biggr)\nonumber\\Ê\nonumber\\
	& &+\frac{\omega^2m_1}{3}\left(\frac{\hbar}{2m_0\omega}\right)^{3/2}\biggl(\hat a^3+\hat a\hat a^+\hat a+\hat a^2\hat a^++\hat a\hat a^{+2} +\hat a^+\hat a^2+\hat a^{+2}\hat a+\hat a^+\hat a\hat a^++\hat a^{+3}\biggr)
\end{eqnarray}
From this expression, we get
\begin{subequations}
\begin{equation}
E_{K,n}^{(1)}=\frac{\sigma}{3}\biggl(\frac{2n^2+2n-1}{4}+\frac{1}{2}\biggr)
\end{equation}
and
\begin{equation}
E_{K,n}^{(2)}=-\frac{1}{\hbar\omega}\left\{(\eta-m_1\beta)^2(3n^2+3n+2)+(3\eta+m_1\beta)^2(3n^2+3n+1)+\left(\frac{\sigma}{12}\right)^2(4n^3+6n^2+14n+6)\right\}.
\end{equation}
\end{subequations}
So, the energy associated to the quantization of the constant of motion in the nth-state is
\begin{eqnarray}\label{energyK}
E_{K,n}&=&\hbar\omega(n+1/2)+\frac{\sigma}{3}\biggl(\frac{2n^2+2n-1}{4}+\frac{1}{2}\biggr)\nonumber\\
& & -\frac{1}{\hbar\omega}\left\{(\eta-m_1\beta)^2(3n^2+3n+2)+(3\eta+m_1\beta)^2(3n^2+3n+1)+\left(\frac{\sigma}{12}\right)^2(4n^3+6n^2+14n+6)\right\}.\nonumber\\
\end{eqnarray}
The difference in the energy levels for the two methods of quantization ($\Delta E_n=E_{H,n}-E_{K,n}$) is 
\begin{equation}\label{diff}
\Delta E_n=\frac{2\sigma}{3}\left(\frac{1}{4}(2n^2+2n-1)+\frac{1}{2}\right)+\frac{4m_1\eta\beta}{\hbar\omega}(6n^2+6n+1)-\frac{\sigma^2}{18\hbar\omega}
(4n^3+6n^2+14n+6). 
\end{equation}
this difference is plotted in the next figure for $m_0=10^{-17}Kg$, $\omega=10~GHz$, and for values of $m_1$ (units of Kg/m) such that $E_{\xi,n}^{(1)}+E_{\xi,n}^{(2)}$ be lower than $1\%$ of $E_n^{(0)}$.
\begin{figure}[H]
\includegraphics[scale=0.70,angle=-90]{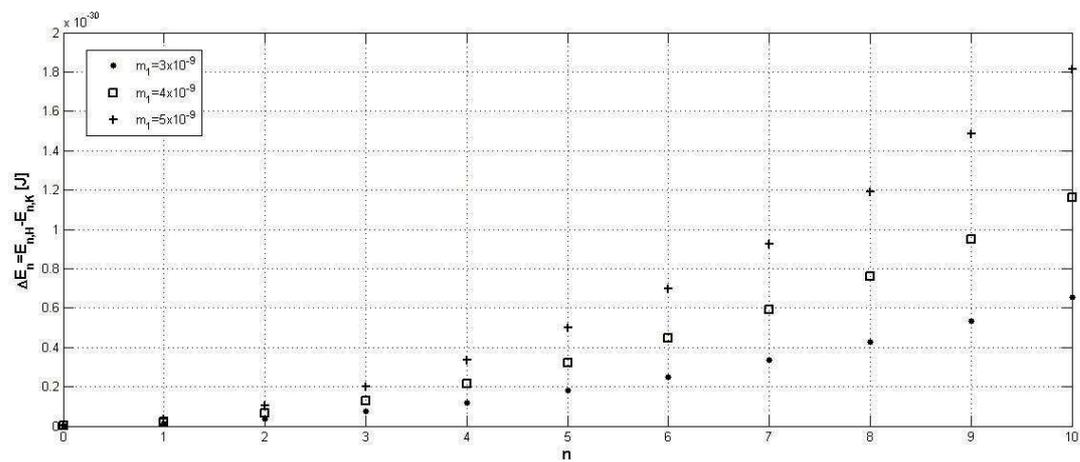}
\centering
    \caption{ Difference on energies for the two cases.  }
\end{figure}
\newpage
\section{ Conclusions}
We have made the study of the quantization of 1-D harmonic oscillator with position depending mass. We used two methods of quantization, one due to the usual Hamiltonian approach, and the other one is proposed approach based on  the quantization of a constant of motion of the system. On both approaches, the Shr\"odinger's equation is used to see the quantum dynamics of the system. The position depending mass produces an additional term which is taken as a perturbation of the usual harmonic oscillator with constant mass. Using perturbation theory at second order, the energies associated to the Hamiltonian and constant of motion approaches was give, obtaining a difference in their values.This difference is not due to the fact that the resulting expression for K is exact , meanwhile for the Hamiltonian is an approximation to second order in Taylor expansion. Finally, it is our expectation that an experiment can be carried out to see whether or not the mass position approach and K-quantization approach can be verified experimentally. 
\newpage\noindent
{\bf References}\\Ê\\
1. H. Goldstein, {\it Classical Mechanics}, Addison-Wesley, (1950).\\ \\
2. S.K. Bose, {\it The rocket problem revisited}, Am. J. Phys., {\bf 51}, (1983), 463.\\ \\
3. A. Sommerfeld, {\it Lectures on Theoretical Physics, vol. I}, Academic Press, (1964).\\ \\
4. H. Gyden, Astronomische Nachrichtem, {\bf 109}, (1884), 1.\\ \\
5, I.V. Meshcherskii, Astronomische Nachrichtem, {\bf 132}, (1893),93.\\ \\
6. C. Tuiega, J. jasiski, T. Iwamoto, and V. Chikan, ACS Nano, {\bf 2}, (2008), 1411.\\ \\
7. H.A. Bethe, Phys. Rev. Lett., {\bf 56}, (1936), 1305.\\ \\
8. E.D. Comminss and P.A. Bucksbaum, {\it Weak Interactions of Leptons and Quarks}, Cambridge University Press, (1983).\\ \\
9. O. Tokamobu, D. Kentare, N. Koichi, and T. Akitomo, Physica  Status Solidi, {\bf B241}, (2004), 2744.\\ \\
10. G.V. L\'opez and E.M. Ju\'arez, J. Mod. Phys., {\bf 4}, (2013), 1638.\\ \\
11. M. Spivak, {\it Physics for Mathematisians, Mechanics I}, Publish or Perish, Inc. (2010).\\ \\
12. G.V. L\'opez and C. Mart\'{\i}nez-Prieto, J. Mod. Phys., {\bf 5}, (2014), 900.\\ \\
13. A. Messiah, {\it Quantum Mechanics}, dover, (1999).\\ \\
14. H. Weyl, {\it Quantummechanick und Gruppentheorie}, Z.T. Physics, {\bf 46}, (1927),1.\\Ê\\
15. S. Cruz y Cruz,  Geometric Methdos in Physics, XXX Workshop 2011. Trends in Mathematics, (2011), 229.\\Ê\\
16. M.S. Cunha and H.R. Christiansen, Comm. Theor. Phys.,{\bf 60}, (2013), 642.\\ \\
17. S. Cruz y Cruz and O. Rosas-Ortiz, J. Phys. A: Math. Theor.,{\bf 42}, (2009), 185205.

\end{document}